\documentclass[11pt, a4paper]{article}
\usepackage[english]{babel}
\usepackage{amsmath}
\usepackage{amssymb,latexsym}
\usepackage[margin=1in]{geometry}

\usepackage{graphicx}
\usepackage{tikz}
\usetikzlibrary{calc}
\usepackage[font=small,labelfont=bf,labelsep=period]{caption}
\usepackage[hang]{subfigure}
\usepackage{url}

\newcommand{\vt}[1]{\ensuremath{\boldsymbol{#1}}} 
\newcommand{\mt}[1]{\ensuremath{\mathsf{#1}}} 
\DeclareMathOperator{\Rel}{Rel}

\begin{document}

\title{Graph measures and network robustness}
\author{W.~Ellens$^{a,b,}$\footnote{Corresponding author.\newline
E-mail addresses: wendy.ellens@tno.nl (W.~Ellens), robert.kooij@tno.nl (R.E.~Kooij).},
R.E.~Kooij$^{a,c}$\\
\footnotesize
$^a$ TNO Information and Communication Technology, P.O. Box 5050, 2600 GB Delft, The Netherlands\\
\footnotesize
$^b$ Mathematical Institute, University of Leiden, P.O. Box 9512, 2300 RA Leiden, The Netherlands\\
\footnotesize
$^{c}$ Faculty of Electrical Engineering, Mathematics and
Computer Science, Delft University of Technology,\\
\footnotesize P.O. Box 5031, 2600 GA
Delft, The Netherlands}
\maketitle

\begin{abstract}
\noindent Network robustness research aims at finding a measure to quantify network robustness. Once such a measure has been established, we will be able to compare networks, to improve existing networks and to design new networks that are able to continue to perform well when it is subject to failures or attacks. In this paper we survey a large amount of robustness measures on simple, undirected and unweighted graphs, in order to offer a tool for network administrators to evaluate and improve the robustness of their network. The measures discussed in this paper are based on the concepts of connectivity (including reliability polynomials), distance, betweenness and clustering. Some other measures are notions from spectral graph theory, more precisely, they are functions of the Laplacian eigenvalues. In addition to surveying these graph measures, the paper also contains a discussion of their functionality as a measure for topological network robustness.
\end{abstract}


\vskip\baselineskip

\section{Introduction}

\subsection{The field of network robustness research}\label{Subsec:Intro}

As we live in a highly networked world, where vital facilities such as hospitals and fire brigades depend on a large amount of networks of different kinds, it is of highest importance that these networks are robust. Think of the consequences if for example telecommunication systems, power grids, water supplies, or road networks are malfunctioning. But what do we mean by network robustness? Let us start by giving a working definition.

\vspace{1\baselineskip}
\emph{Robustness} is the ability of a network to continue performing well when it is subject to failures or attacks.
\vspace{1\baselineskip}

In order to decide whether a given network is robust, a way to quantitatively measure network robustness is needed. Intuitively robustness is all about back-up possibilities \cite{Singer}, or alternative paths \cite{Wu}, but it is a challenge to capture these concepts in a mathematical formula. During the past years a lot of robustness measures have been proposed \cite{Sydney}. This paper aims at giving an overview of the most used measures. Besides this, we evaluate the surveyed measures by assessing them based on the following criteria. In our opinion networks become more robust when links are added, and a connection between two nodes is more robust when there is more than one path between them. Furthermore ---~in order to be used in practice~--- we think it is important that the meaning of a measure is intuitively clear.

Network robustness research is carried out by scientists with different backgrounds, like mathematics, physics, computer science and biology \cite{Singer}. As a result, quite a lot of different approaches to capture the robustness properties of a network have been undertaken \cite{Mieghem2}. All of these approached are based on the analysis of the underlying \emph{graph} ---~consisting of a set of \emph{vertices} connected by \emph{edges}~--- of a network. We will use the words vertices and edges used in graph theory instead of the words nodes and links as these concepts are usually called in network theory. In this paper, unless differently stated, by a graph $G=(V,E)$ we mean a simple, undirected, connected, unweighted, finite and deterministic graph, with $|V|=n$ vertices and $|E|=m$ edges. 

In the field of complex networks a large amount of \emph{graph measures} (also called \emph{graph metrics} or \emph{graph invariants}) have been studied. For a review of these measures see for example \cite{Boccaletti,Costa,Dorogovtsev}. We focus on these measures that have been proposed for, or are intuitively relevant for, evaluating the robustness of a network. The graph measures considered in this paper are topological measures, indicating that they describe the \emph{network topology} (the geografical design consisting of vertices connected by edges), neglecting any processes running on top of the network.

\subsection{Outline}

The rest of this paper is divided into three main sections. The first section (Section \ref{Sec:ClassicalMeasures}) contains a review of some classical graph measures. Subsections \ref{Subsec:Connectivity} until \ref{Subsec:Clustering} consider a broad range of classical graph measures from complex network theory, such as vertex and edge connectivity, graph diameter, average vertex betweenness and clustering coefficient. The central question is whether these measures, which are not specifically introduced as network robustness measures, could be used to determine the robustness properties of a graph. The subject of Subsection \ref{Subsec:Relpol} is the reliability polynomial, which strictly speaking is not a graph measure, since it gives a function instead of a single number for a given graph, but it represents a classical method to measure network robustness. 

The second main section is Section \ref{Sec:SpectralMeasures}. All three measures discussed in this section have specifically been proposed as network robustness measures and all three of them are based on the Laplacian spectrum. Subsection \ref{Subsec:AlgConn} is about the second smallest Laplacian eigenvalue, the algebraic connectivity. The measures treated in Subsections \ref{Subsec:SpanningTrees} and \ref{Subsec:EffResist} are respectively the number of spanning trees and the effective graph resistance. Both measures are based on the complete spectrum of the Laplacian.

Section \ref{Sec:Evaluation} is the third main section. It does not contain any new graph measures, but evaluates the fourteen measures introduced in Section \ref{Sec:ClassicalMeasures} and Section \ref{Sec:SpectralMeasures}. The evaluation section assesses the robustness measures by means of a selection of small example graphs. Furthermore, it checks whether the measures satisfy the criteria stated in Subsection \ref{Subsec:Intro}; a robustness measure should be able to detect the addition of an edge and it should consider the back-up possibilities in a graph. At last, it should be intuitively clear that the measure indeed captures the robustness of a graph. 

These three main sections are followed by a conclusion section (Section \ref{Sec:Conclusion}), which recapitulates the findings of earlier sections and suggests a direction for further research in the field of network robustness.

\section{Classical graph measures}\label{Sec:ClassicalMeasures}

In the past decades, numerous measures have been introduced to characterise graphs. In this section we treat these classical graph measures that are intuitively relevant for evaluating the robustness of a network. Each subsection describes and discusses a specific graph measure or a class of related measures. Subsection \ref{Subsec:Connectivity} is about graph connectivity, vertex connectivity and edge connectivity. Subsection \ref{Subsec:Distance} discusses these measures based on distance (path length in number of edges) in a graph; average vertex distance, graph diameter and graph efficiency. The concept of betweenness ---~covering the measures average vertex betweenness, average edge betweenness and maximum edge betweenness~--- is the subject of Subsection \ref{Subsec:Betweenness}. Subsection \ref{Subsec:Clustering} treats the clustering coefficient and Subsection \ref{Subsec:Relpol} is about the reliability polynomial of a graph.

\subsection{Connectivity}\label{Subsec:Connectivity}

Apart from the classical binary \emph{connectivity} measure $\kappa$, which distinguishes \emph{connected} graphs ($\kappa=1$) having paths between all pairs of vertices and \emph{unconnected} graphs ($\kappa=0$) for which at least one pair of vertices lacks a connecting path, two more connectivity measures have been defined: vertex and edge connectivity \cite{Diestel}. 

The \emph{vertex connectivity} $\kappa_v$ of an incomplete graph is the minimal number of vertices to be removed in order to disconnect it. The number of edges that need to be removed to disconnect the graph is called the \emph{edge connectivity} $\kappa_e$. It is easy to see that $\kappa_v\leq\kappa_e\leq\delta_{\min}$ \cite{Diestel}, where $\delta_{\min}$ is the minimum degree of the vertices. For a complete graph $K_n$ the vertex connectivity cannot be determined by the definition above, because it cannot be disconnected by deleting vertices. In order for the inequality $\kappa_v\leq\kappa_e\leq\delta_{\min}$ to hold also in the case of a complete graph, its vertex connectivity is defined to be $\kappa_v=n-1$. It seems natural to say that the higher the vertex or edge connectivity of a graph, the more robust it is.

\subsection{Distance}\label{Subsec:Distance}
Let the \emph{distance} $d_{ij}$ be the length (number of edges) of the shortest path between vertices $i$ and $j$. The maximum $d_{\max}$ over these distances is called the \emph{diameter} and the average over all pairs is denoted by $\bar{d}$, 

	\begin{equation}\bar{d}=\frac{2}{n(n-1)}\sum_{i=1}^n\sum_{j=i+1}^n d_{ij}.\end{equation}
The average distance is equal to $\frac{2}{n(n-1)}$ times the Wiener index \cite{Wiener} (the sum of the lengths of the shortest paths). The meaning of the diameter and the average distance as robustness measures follows from the fact that the shorter a path, the robuster it is. Nevertheless, a vulnerable path can be compensated by adding back-up paths, which are not considered by the two measures, this clearly is a disadvantage. The average distance is more sensible than the diameter, as the first is strictly decreasing when edges are added, while the latter may remain equal while adding edges.

Another measure based on the notion of distance in a graph is the \emph{efficiency}, denoted $E$ \cite{Latora}.

	\begin{equation}E=\frac{2}{n(n-1)}\sum_{i=1}^n\sum_{j=i+1}^n \frac{1}{d_{ij}}.\end{equation}
For the efficiency it holds that the greater the value, the greater the robustness, because the reciprocals of the path lengths are used. The advantage of this measure is that it can be used for unconnected networks, such as social networks or networks subject to failures. Otherwise, it has the same disadvantage as the average path length; alternative paths are not considered.

\subsection{Betweenness}\label{Subsec:Betweenness}

The \emph{betweenness} denotes the number of shortest paths between pairs of vertices, passing through a vertex or an edge $x$. If there exists more than one shortest path between two vertices, then each of these $k$ paths is counted $1/k$ times. The formal definition of the betweenness of a vertex or an edge $x$ is
$$b_x=\sum_{i=1}^n\sum_{j=i+1}^n \frac{n_{ij}(x)}{n_{ij}},$$
where $n_{ij}(x)$ is the number of shortest paths between $i$ and $j$ passing through $x$ and $n_{ij}$ is the total number of shortest paths between $i$ and $j$. The vertex betweenness is sometimes called betweenness centrality, because it has been introduced to determine the vertices that occupy central positions in the network \cite{Freeman}. 

The reason why we have included betweenness in this survey of robustness measures is as follows. Suppose there is one unit of traffic between all pairs of vertices and traffic travels by shortest paths (dividing the load if there is more than one shortest path), then the load of a vertex/edge is given by its betweenness. Deleting vertices or edges with a higher load can have more impact than deleting others. Betweenness can therefore help to identify bottlenecks and give a tool to improve the robustness of a network. However, the existence of alternative paths for network elements with a high load is not considered. Like distance, betweenness is thus a measure based on shortest paths only.

In order to get a measure for the robustness of a network we can take the average of the vertex/edge betweenness. The smaller this average, the more robust the network. It turns out that the average vertex $\bar{b}_v$ and edge betweenness $\bar{b}_e$ are linear functions of the average distance. See \cite{Ellens2} for the derivation. 

	\begin{align*}
		\bar{b}_v& =\frac{1}{2}(n-1)(\bar{d}+1),\\
		\bar{b}_e& =\frac{n(n-1)}{2m}\bar{d}.\\
	\end{align*}
As a consequence of these linear relations, the average distance and the average vertex betweenness will always indicate the same graph as most robust when comparing the robustness of two graphs, provided the graphs have the same number of vertices. The same holds for the three measures (average distance, average vertex betweenness and average edge betweenness) when the number of vertices and edges of the graphs are equal.

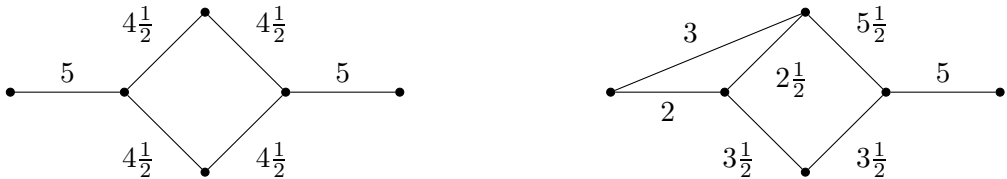
\begin{figure}[h,t]
\begin{center}

\subfigure[Graph with maximum edge betweenness of 5\label{fig:betweenness1}]{
\begin{minipage}{.48\textwidth}
\begin{center}
\begin{tikzpicture}[scale=1.5,rotate=90,
	mynodes/.style={draw,circle,fill=black,minimum size=3pt,inner sep=0pt}]
	
 	\draw (0,0) node(1)[mynodes]{} 
	 	++(0,-1) node(2)[mynodes]{} 
		++(-135:1) node(4)[mynodes]{} 
		++(-45:1) node(5)[mynodes]{} 
		++(0,-1) node(6)[mynodes]{}
		(2)+(-45:1) node(3)[mynodes]{};
	
	\draw 		
		(1) to node[auto]{$5$} (2)
		(2) to node[auto]{$4\frac{1}{2}$} (3) 
		(2) to node[auto, swap]{$4\frac{1}{2}$} (4) 
		(3) to node[auto]{$4\frac{1}{2}$} (5) 
		(4) to node[auto, swap]{$4\frac{1}{2}$} (5) 
		(5) to node[auto]{$5$} (6);
\end{tikzpicture}
\end{center}
\vskip1\baselineskip
\end{minipage}}
\subfigure[Graph with maximum edge betweenness of $5\frac{1}{2}$\label{fig:betweenness2}]{
\begin{minipage}{.48\textwidth}
\begin{center}
\begin{tikzpicture}[scale=1.5,rotate=90,
	mynodes/.style={draw,circle,fill=black,minimum size=3pt,inner sep=0pt}]
	
 	\draw (0,0) node(1)[mynodes]{} 
	 	++(0,-1) node(2)[mynodes]{} 
		++(-135:1) node(4)[mynodes]{} 
		++(-45:1) node(5)[mynodes]{} 
		++(0,-1) node(6)[mynodes]{}
		(2)+(-45:1) node(3)[mynodes]{};
	
	\draw 
		(1) to node[auto, swap]{$2$} (2)
		(2) to node[auto, swap]{$2\frac{1}{2}$} (3) 
		(2) to node[auto, swap]{$3\frac{1}{2}$} (4) 
		(3) to node[auto]{$5\frac{1}{2}$} (5) 
		(4) to node[auto, swap]{$3\frac{1}{2}$} (5) 
		(5) to node[auto]{$5$} (6)	
		(1) to node[auto]{$3$} (3);
\end{tikzpicture}
\end{center}
\vskip1\baselineskip
\end{minipage}}

\end{center}
\caption{The maximum edge betweenness can increase when an edge is added. The betweenness of each edge is given in the graphs.}
\label{fig:betweenness}
\end{figure}

Sydney et al. have proposed a robustness measure based on the maximum edge betweenness $b_e^{\max}$ and its behaviour as vertices are removed, because this maximum determines the bandwidth that can be assigned to each flow \cite{Sydney}. The maximum edge betweenness has a problem though; it can increase while an edge is added, while we believe that the network becomes more robust when edges are added. We give an example in Figure \ref{fig:betweenness}.

\subsection{Clustering}\label{Subsec:Clustering}

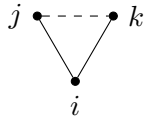
\begin{figure}[h,t]
\begin{center}

\begin{tikzpicture}[scale=1,
	mynodes/.style={draw,circle,fill=black,minimum size=3pt,inner sep=0pt}]
	
 	\draw (0,0) node(1)[mynodes,label=below:$i$]{} 
		+(120:1) node(2)[mynodes,label=left:$j$]{} 
		+(60:1) node(3)[mynodes,label=right:$k$]{};
	
	\draw (1) -- (2) 
		(1) -- (3);
	\draw[dashed]
		(2) -- (3);
\end{tikzpicture}

\end{center}
\caption{Vertices $j,k$ sharing a neighbour $i$ may or may not be neighbours themselves.}
\label{fig:clustering}
\end{figure}

The presence of triangles is captured by the \emph{clustering coefficient} \cite{Watts}, which compares the number of triangles to the number of connected triples. The clustering coefficient gives the portion of vertices $j,k$ sharing a neighbour $i$ that are also neighbours themselves (which means that the edge $(j,k)$ is present, see Figure \ref{fig:clustering}). The clustering coefficient $c_i$ of a vertex $i$ is defined as the number of edges among neighbours of $i$ divided by $\delta_i(\delta_i-1)/2$, the total possible number of edges among its neighbours. Here $\delta_i$ is the degree (number of neighbours) of a vertex $i$. The overall clustering coefficient of a graph is the average over the clustering coefficients of the vertices. This definition gives

	\begin{multline*}C=\frac{1}{n}\sum_{i\in V;\delta_i>1}c_i =\frac{1}{n}\sum_{i\in V;\delta_i>1} \frac{2}{\delta_i(\delta_i-1)}e_i =\\
	\frac{1}{n}\sum_{i\in V;\delta_i>1} \frac{1}{\delta_i(\delta_i-1)}\sum_{j=1}^n\sum_{k=1}^n a_{ij}a_{jk}a_{ki} =\frac{1}{n}\sum_{i\in V;\delta_i>1} \frac{1}{\delta_i(\delta_i-1)}\left(\mt{A}^3\right)_{ii},\end{multline*}
with $e_v$ the number of edges among neighbours of $v$, and $a_{ij}$ the $ij$-th element of the \emph{adjacency matrix} $\mt{A}$, which is equal to one if the edge $(i,j)$ is present and zero otherwise.

Although the clustering coefficient was originally designed for social networks, in which it measures the probability that two friends of a person are friends of each other too, it can also be used to measure robustness in other types of networks. A high clustering coefficient indicates high robustness, because the number of alternative paths grows with the number of triangles.

\subsection{Reliability polynomials}\label{Subsec:Relpol}

Although the reliability polynomial is not part of the standard set of graph measures, we treat it in this chapter, because it is a classical way to quantify network robustness. Reliability polynomials are based on the notion of graph connectivity. However, we dedicate a new subsection to reliability polynomials, because they are derived by a probabilistic approach, unlike the classical connectivity measures discussed in Subsection \ref{Subsec:Connectivity}. The \emph{reliability polynomial} \cite{Moore} $\Rel(G)$ of a graph $G$ is equal to the probability that the graph is connected when each edge is (independently of the others) present with probability $p=1-q$, in other words
		$$\Rel(G)=\sum_{i=0}^m F_i(1-p)^ip^{m-i},$$
when $F_i$ denotes the number of sets of $i$ edges whose removal leaves $G$ connected.

Reliability polynomials are an intuitive way to measure network robustness, although it is difficult to decide what value we should assign to $p$. The robustness evaluation of graphs depends on the value of $p$; pairs of graphs for which the reliability polynomial of the first graph is larger for small $p$, while the reliability polynomial of the second is larger for large $p$, are known \cite{Kelmans}. It seems reasonable to consider $p$ to be close to one, because in real-world networks edge failures are scarce. 

It has been stated in \cite{Moore} that the reliability polynomial for $p$ close to one always give the same evaluation on robustness as the edge connectivity. More precisely, the relation between the reliability polynomial $\Rel(G)$ of a graph $G$ and the edge connectivity $\kappa_e(G)$ satisfies the following two properties
	\begin{enumerate}
		\item If $\kappa_e(G_1)<\kappa_e(G_2)$, then for $p$ close enough to one we have $\Rel(G_1)<\Rel(G_2)$. This means that the reliability polynomial for $p$ close to one and the edge connectivity give the same evaluation on network robustness.
		\item Let $s(G)$ be the number of subsets of $\kappa_e(G)$ edges whose removal disconnects $G$. If $\kappa_e(G_1)=\kappa_e(G_2)$ and $s(G_1)>s(G_2)$ then for $p$ close enough to one we have $\Rel(G_1)<\Rel(G_2)$. 
	\end{enumerate}	

A proof can be found in \cite{Ellens2}. Remark that a reliability polynomial can also be defined for vertex deletion instead of edge deletion. In that case the reliability polynomial for $p$ close to one and the vertex connectivity give the same robustness evaluation.

\section{Spectral graph measures}\label{Sec:SpectralMeasures}

Networks can be represented by graphs. These graphs can be studied directly, as we have done in the previous chapters, but also by looking at the matrices associated to a graph. One of these matrices is the \emph{Laplacian}. The Laplacian \mt{L} is the difference $\mt{\Delta}-\mt{A}$ of the degree matrix \mt{\Delta} and the adjacency matrix \mt{A}, i.e.
		$$\mt{L}_{ij}=\begin{cases}
										\delta_i 	&\text{if }i=j\\
										-1 				&\text{if }(i,j)\in E\\
										0					&\text{otherwise}
									\end{cases}.$$
For more information we refer to \cite{Mohar, Ellens2}. Several robustness measures based on the eigenvalues of the Laplacian have been proposed. We treat three of those measures; the algebraic connectivity in Subsection \ref{Subsec:AlgConn}, the number of spanning trees in Subsection \ref{Subsec:SpanningTrees} and the effective graph resistance in Subsection \ref{Subsec:EffResist}.

\subsection{Algebraic connectivity}\label{Subsec:AlgConn}

Because the Laplacian is symmetric, positive semidefinite and the rows sum up to 0, its eigenvalues are real, non-negative and the smallest one is zero. Hence, we can order the eigenvalues and denote them as $\lambda_i$ for $i=1,\ldots,n=|V|$ such that $0=\lambda_1\leq\lambda_2\leq\cdots\leq\lambda_n$. We denote vector with elements $\lambda_i$ by $\vt{\lambda}$. The second smallest eigenvalue $\lambda_2$ of the Laplacian is called \emph{algebraic connectivity} by Miroslav Fiedler \cite{Fiedler}. There are a few reasons to believe that it is a measure for the connectivity of a graph:

\begin{enumerate}
	\item The algebraic connectivity is equal to zero if and only if the graph is unconnected.
	\item The algebraic connectivity of an incomplete graph is not greater than the vertex connectivity. Therefore we have:

	$$0\leq\lambda_2\leq\kappa_v\leq\kappa_e\leq\delta_{\min}.$$
\end{enumerate}

\begin{figure}[ht]
\begin{center}

\subfigure[$\vt{\lambda}=(0,2,2,4)$\label{fig:algconn1}]{
\begin{minipage}[b]{.2\textwidth}
\begin{center}
\begin{tikzpicture}[scale=1,
	mynodes/.style={draw,circle,fill=black,minimum size=3pt,inner sep=0pt}]
	
 	\draw (0,0) node(1)[mynodes]{} 
	 	++(1,0) node(2)[mynodes]{} 
		++(0,-1) node(3)[mynodes]{} 
		++(-1,0) node(4)[mynodes]{} ;
	
	\draw (1)--(2) 
		(1)--(4)
		(2)--(3) 
		(3)--(4);
\end{tikzpicture}
\end{center}
\vskip1\baselineskip
\end{minipage}}
\subfigure[$\vt{\lambda}=(0,2,4,4)$\label{fig:algconn2}]{
\begin{minipage}[b]{.2\textwidth}
\begin{center}
\begin{tikzpicture}[scale=1,
	mynodes/.style={draw,circle,fill=black,minimum size=3pt,inner sep=0pt}]
	
 	\draw (0,0) node(1)[mynodes]{} 
	 	++(1,0) node(2)[mynodes]{} 
		++(0,-1) node(3)[mynodes]{} 
		++(-1,0) node(4)[mynodes]{} ;
	
	\draw (1)--(2) 
		(1)--(3)
		(1)--(4)
		(2)--(3) 
		(3)--(4);
\end{tikzpicture}
\end{center}
\vskip1\baselineskip
\end{minipage}}
\end{center}
\caption{Two graphs with identical algebraic connectivity}
\label{fig:algconn}
\end{figure}
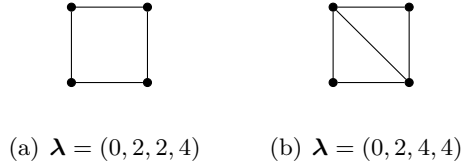

Beside the fact that it is not intuitively clear which properties of the graph the algebraic connectivity expresses, as a measure for network robustness it also has the problem that is not strictly increasing when an edge is added. Figure \ref{fig:algconn} shows the example of \cite{Baras}. In order to guarantee that a measure strictly increases when adding edges, it is not enough to base the measure on the first (fixed number) $k$ Laplacian eigenvalues \cite{Ellens2}, therefore the measures in the following subsections are a function of the whole Laplacian spectrum.

\subsection{Number of spanning trees}\label{Subsec:SpanningTrees}

Baras and Hovareshti suggest the number of spanning trees (a \emph{spanning tree}\index{Spanning tree} is a subgraph containing $n-1$ edges and no cycles) as an indicator of network robustness \cite{Baras}. It is a consequence of Kirchhoff's matrix-tree theorem that the number of spanning trees $\xi$ can be written as a function of the unweighted Laplacian eigenvalues:

		$$\xi=\frac{1}{n}\prod_{i=2}^n\lambda_i.$$
See \cite{Ellens2} for a rigorous proof.

The number of spanning trees gives the same judgment about the robustness of a network as the reliability polynomial gives when $p$ goes to zero \cite{Colbourn}. In other words, if $\xi(G_1)<\xi(G_2)$, then for $p$ close enough to zero we have $\Rel(G_1)<\Rel(G_2)$, a proof of which can be found in \cite{Ellens2}.

\subsection{Effective resistance}\label{Subsec:EffResist}

Assume the graph is seen as an electrical circuit, where an edge $(i,j)$ corresponds to a resistor of $r_{ij}=1$ Ohm. Informally, the \emph{effective resistance} between two vertices of a network ---~the resistance of the total system when a voltage is connected across them~--- can be calculated by the well-known series and parallel manipulations. Two edges, corresponding to resistors with resistance $r_1=1$ and $r_2=1$ Ohm, in series can be replaced by one edge with effective resistance $r_1+r_2=1+1=2$ Ohm. If the two edges are connected in parallel, then they can be replaced by an edge with effective resistance $\left(r_1^{-1}+r_2^{-1}\right)^{-1}=\left(1^{-1}+1^{-1}\right)^{-1}=1/2$ Ohm. The \emph{effective graph resistance} is the sum of the effective resistances over all pairs of vertices.

More formally, for each pair of vertices the effective resistance between these vertices can be calculated by Kirchhoff's circuit laws. Let a voltage be connected between vertices $a$ and $b$ and let $I>0$ be the net current out of source $a$ and into sink $b$, \emph{Kirchhoff's current law} states that the current $y_{ij}$ between vertices $i$ and $j$ (where $y_{ij}=-y_{ji}$) must satisfy
	\begin{equation}\label{eq:Kirchhoff}
		\sum_{j\in N(i)}y_{ij}=
			\begin{cases}
				I 	&\text{if }y=a\\
				-I 	&\text{if }y=b\\
				0		&\text{otherwise},
			\end{cases}
	\end{equation}	
with $N(i)$ the neighbourhood of $i$, that is, the set of vertices adjacent to vertex $i$. This first law means that the total flow into a vertex equals the total flow out of it. The second of Kirchhoff's laws, \emph{Kirchhoff's voltage law}, is equivalent to saying that a potential $v$ may be associated with any vertex $i$, such that for all edges $(i,j)$
	\begin{equation}\label{eq:Ohm}y_{ij}r_{ij}=v_i-v_j.\end{equation}
This is called \emph{Ohm's law}. The \emph{effective resistance} $R_{ab}$ between vertices $a$ and $b$ is uniquely \cite{KleinRandic} defined as
	$$R_{ab}=\frac{v_a-v_b}{I}.$$
The \emph{effective graph resistance} $R$, also called \emph{total effective resistance} or \emph{Kirchhoff index}, is defined as the sum of the effective resistances over all pairs of vertices. Klein and Randi\'c \cite{KleinRandic} have proved that it can be written as a function of the non-zero Laplacian eigenvalues:
	$$R =\sum_{1\leq i<j\leq n}R_{ij}=n\sum_{i=2}^n\frac{1}{\lambda_i}.$$
For more information on the properties of the pairwise effective resistance and the total effective resistance see \cite{KleinRandic, Ellens}.

In \cite{Ellens} it has been proven that the effective graph resistance strictly decreases when an edge is added. In addition, another argument to adopt the effective graph resistance as a measure for network robustness have been given in the same paper. The pairwise effective resistance gives the vulnerability of a connection between a pair of vertices that takes into account both the number of paths between the vertices and their length, considering the number of back-up paths as well as their quality. A small value of the effective graph resistance therefore indicates a robust network.

\section{An evaluation of robustness measures}\label{Sec:Evaluation}

In this section we compare the graph measures described in Section \ref{Sec:ClassicalMeasures} and Section \ref{Sec:SpectralMeasures} and evaluate their ability to capture the robustness properties of a network. We start (in Subsection \ref{Subsec:Examples}) by a comparison by means of some small example graphs and continue (in Subsection \ref{Subsec:Criteria} by verifying whether the measures meet the criteria mentioned in Section \ref{Subsec:Intro}.

\subsection{Examples}\label{Subsec:Examples}

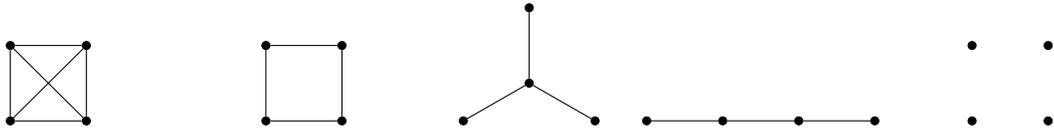
\begin{figure}[!thbp]
\begin{center}

\subfigure[Complete graph $K_4$\label{fig:measures1}]{
\begin{minipage}[b]{.2\textwidth}
\begin{center}
\begin{tikzpicture}[scale=1,
	mynodes/.style={draw,circle,fill=black,minimum size=3pt,inner sep=0pt}]
	
 	\draw (0,0) node(1)[mynodes]{} 
	 	++(1,0) node(2)[mynodes]{} 
		++(0,-1) node(3)[mynodes]{} 
		++(-1,0) node(4)[mynodes]{} ;
	
	\draw (1)--(2) 
		(1)--(3)
		(1)--(4)
		(2)--(3) 
		(2)--(4) 
		(3)--(4);
\end{tikzpicture}
\end{center}
\vskip1\baselineskip
\end{minipage}}
\subfigure[Cycle graph $C_4$\label{fig:measures2}]{
\begin{minipage}[b]{.19\textwidth}
\begin{center}
\begin{tikzpicture}[scale=1,
	mynodes/.style={draw,circle,fill=black,minimum size=3pt,inner sep=0pt}]
	
 	\draw (0,0) node(1)[mynodes]{} 
	 	++(1,0) node(2)[mynodes]{} 
		++(0,-1) node(3)[mynodes]{} 
		++(-1,0) node(4)[mynodes]{} ;
	
	\draw (1)--(2) 
		(1)--(4)
		(2)--(3) 
		(3)--(4);
\end{tikzpicture}
\end{center}
\vskip1\baselineskip
\end{minipage}}
\subfigure[Star graph $S_4$\label{fig:measures3}]{
\begin{minipage}[b]{.15\textwidth}
\begin{center}
\begin{tikzpicture}[scale=1,
	mynodes/.style={draw,circle,fill=black,minimum size=3pt,inner sep=0pt}]
	
 	\draw (0,0) node(1)[mynodes]{} 
	 	++(90:1) node(2)[mynodes]{} 
		(1)+(-150:1) node(3)[mynodes]{} 
		(1)+(-30:1) node(4)[mynodes]{} ;
	
	\draw (1)--(2) 
		(1)--(3)
		(1)--(4);
\end{tikzpicture}
\end{center}
\vskip1\baselineskip
\end{minipage}}
\subfigure[Path graph $P_4$\label{fig:measures4}]{
\begin{minipage}[b]{.2\textwidth}
\begin{center}
\begin{tikzpicture}[scale=1,
	mynodes/.style={draw,circle,fill=black,minimum size=3pt,inner sep=0pt}]
	
 	\draw (0,0) node(1)[mynodes]{} 
	 	++(1,0) node(2)[mynodes]{} 
		++(1,0) node(3)[mynodes]{} 
		++(1,0) node(4)[mynodes]{} ;
	
	\draw (1)--(2) 
		(2)--(3) 
		(3)--(4);
\end{tikzpicture}
\end{center}
\vskip1\baselineskip
\end{minipage}}
\subfigure[Empty graph $O_4$\label{fig:measures5}]{
\begin{minipage}[b]{.18\textwidth}
\begin{center}
\begin{tikzpicture}[scale=1,
	mynodes/.style={draw,circle,fill=black,minimum size=3pt,inner sep=0pt}]
	
 	\draw (0,0) node(1)[mynodes]{} 
	 	++(1,0) node(2)[mynodes]{} 
		++(0,-1) node(3)[mynodes]{} 
		++(-1,0) node(4)[mynodes]{} ;
\end{tikzpicture}
\end{center}
\vskip1\baselineskip
\end{minipage}}

\end{center}
\caption{Examples of graphs with four vertices}
\label{fig:measures}
\end{figure}

\begin{figure}[!htbp]
\begin{center}

\begin{tikzpicture}[scale=4]
\draw[->] (0,-0.1) -- (1,-0.1) node[below] {1};
\draw[->] (0,-0.1) -- (0,1) node[left] {1};
\draw (-0.02,0.5) node[left] {$\Rel$};
\draw (0.5,-0.12) node[below] {$p$};
\draw (0,0) node[left] {0};
\draw (0,-0.1) node[below] {0};
\draw[domain=0:1] plot (\x,-6*\x^6 + 24*\x^5 - 33*\x^4 + 16*\x^3) + (-0.4,-0.05) node[right=5pt,fill=white] {$K_4$} ;
\draw[densely dashed][domain=0:1] plot (\x,-3*\x^4+4*\x^3) +(-0.4,-0.3) node[right=5pt,fill=white] {$C_4$};
\draw[dotted][domain=0:1] plot (\x,\x^3) +(-0.4,-0.6) node[right=5pt,fill=white] {$S_4$/$P_4$};
\draw[dashed][domain=0:1] plot (\x,0) +(-0.4,0) node[right=5pt,fill=white] {$O_4$};
\end{tikzpicture}

\vskip\baselineskip
\caption{Graphs of the reliability polynomials for the five graphs of Figure \ref{fig:measures}}
\label{fig:graphsrelpols}
\end{center}
\end{figure}
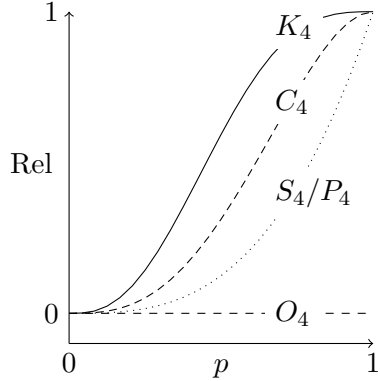

We start by calculating the values of all measures for the example graphs with four vertices depicted in Figure \ref{fig:measures}. The results are given in Table \ref{tab:measures} and Figure \ref{fig:graphsrelpols}.

\begin{table}[!ht]
	\begin{center}
	\begin{tabular}{l|*{13}{l}}
	&$\kappa$ &$\kappa_v$ &$\kappa_e$ &$d_{\max}$ &$\bar{d}$ &$E$	&$b_e^{\max}$ &$\bar{b}_v$ &$\bar{b}_e$ &$C$	&$\lambda_2$	&$\xi$	&R\\
	\hline
	$K_4$	&1	&3	&3	&1				&1							&1							&1	&3							&1							&1	&4		&16	&3\\
	$C_4$	&1	&2	&2	&2				&$1\frac{1}{3}$	&$\frac{5}{6}$	&2	&2							&2							&0	&2		&4	&5\\
	$S_4$	&1	&1	&1	&2				&$1\frac{1}{2}$	&$\frac{3}{4}$	&3	&$2\frac{1}{4}$	&3							&0	&1		&1	&9\\
	$P_4$	&1	&1	&1	&3				&$1\frac{2}{3}$	&$\frac{13}{18}$&4	&$2\frac{1}{2}$	&$3\frac{1}{3}$	&0	&0.59	&1	&10\\
	$O_4$	&0	&0	&0	&$\infty$	&$\infty$				&0							&-	&-							&-							&0	&0		&0	&$\infty$\\
	\end{tabular}
	\end{center}
	\caption{The values of some graph measures for the five graphs of Figure \ref{fig:measures}. The measure are: connectivity, vertex connectivity, edge connectivity, diameter, average distance, graph efficiency, maximum edge betweenness, average vertex betweenness, average edge betweenness, clustering coefficient, algebraic connectivity, number of spanning trees and effective graph resistance.}\label{tab:measures}
\end{table}

Our intuition says that the graphs are ordered by decreasing robustness. The robustness evaluations of the measures of Table \ref{tab:measures} and Figure \ref{fig:graphsrelpols} correspond to this intuition. Although not all of them can distinguish all graphs, all of the measures would say that the graphs are indeed in order of decreasing robustness.

By analysing the values in the table we come to the conclusion that the clustering coefficient and the connectedness are poor robustness measures, because they have the same value for four out of five graphs. Also the vertex and edge connectivity, the diameter, the reliability polynomial and the number of spanning trees cannot distinguish all graphs. More specifically, the connectivity measures (including the reliability polynomial) and the number of spanning trees are constant on the set of trees. The only measures that can distinguish unconnected graphs are the graph efficiency and the clustering coefficient.

The maximum edge betweenness performs well in this example, but has been proved to fail in other situations like that of Figure \ref{fig:betweenness}. Also the algebraic connectivity distinguishes the example graphs. However, Figure \ref{fig:algconn} has shown that it does not always detect the addition of an edge. The average distance, the average vertex betweenness and the average edge betweenness ---~which have been shown to always classify graphs (with a given number of vertices) in the same order~--- seem to be good robustness measures. 

Nevertheless, the disadvantage of these measures is that they consider only the shortest paths in a graph while for the robustness of a network also the (longer) alternative paths are important. The same holds for the graph efficiency. The only measure that gives the desired evaluation for the example graphs and also measures back-up paths in the graph, is the effective graph resistance.

\subsection{Criteria control}\label{Subsec:Criteria}

We now evaluate the presented robustness measures by checking the criteria of Subsection \ref{Subsec:Intro}; does the measure increase at the addition of edges, are alternative paths taken into account, is the measure intuitive enough? The requirement that a robustness measure must be strictly increasing when an edge is added, excludes a lot of the measures mentioned above, except for the average distance (and the average vertex/edge betweenness which are related to the average distance), the graph efficiency, the reliability polynomial, the number of spanning trees and the effective graph resistance. 

The reliability polynomial is a function of $p$, the failure probability of an edge being $1-p$. For `$p$ close to zero', the reliability polynomial and the number of spanning trees are related and give the probability that the graph is connected when an edge is removed with large probability, which is a graph property that is not compatible with the fact that failures in real-world networks are scarce. Therefore the reliability polynomial `for $p$ close to one' seems a better measure. Nevertheless, this case corresponds to the edge connectivity, which does not strictly increase when edges are added. 

Since the average distance and efficiency measure the length of the average (inverse) shortest path between a pair of vertices and do not take the number and length of alternative paths into account, the effective graph resistance seems to best capture the robustness properties. However, the meaning of the effective graph resistance with respect to network robustness is not directly clear, which is a disadvantage in practical situations.

\section{Conclusion}\label{Sec:Conclusion}

Network robustness research aims at finding a measure to quantify network robustness. Once such a measure has been established, we will be able to compare networks, to improve existing networks and to design robust new networks. This paper has given an overview measures on simple, undirected and unweighted graphs that are or can be used for measuring network robustness. Eleven measures discussed in this paper are based on the concepts of connectivity (including reliability polynomials), distance, betweenness, or clustering. Three more measures are notions from spectral graph theory, more precisely, they are functions of the Laplacian eigenvalues. In addition to surveying these graph measures, the paper also contains a discussion of their functionality as a robustness measure.

The analysis of the fourteen measures have shown that all measures are able to place some small example graphs in the same order of robustness as we would intuitively place them. Not all measures are able to distinguish the given graphs; this holds for the four connectivity measures (the reliability polynomial included), the clustering coefficient and the number of spanning trees. All connectivity measures, the diameter, the clustering coefficient and the algebraic connectivity may stay equal when an edge is added, while the network is believed to become more robust in this case. The maximum edge betweenness may even increase ~---and thus indicate that the network becomes less robust~--- at the addition of an edge. The only measures that can be used for networks consisting of several unconnected components ---~such as social networks or networks under attack~--- are the graph efficiency and the clustering coefficient.

The reliability polynomial is a realistic measure for very small failure probabilities, but in this case it is related to the edge connectivity, which is not a very sensitive measure. For large failure probabilities it is related to the number of spanning trees, which is a disadvantage for the number of spanning trees as a robustness measure. The algebraic connectivity and the effective graph resistance have the disadvantage that it is not easy to explain their meaning to network administrators. However, the effective graph resistance is the only measure that considers the back-up possibilities in a graph.

We have surveyed a large amount of robustness measures on simple, undirected and unweighted graphs, with the aim of offering a tool for network administrators to evaluate and improve the robustness of their network. In practice, not all edges in a graph are equally important. Some of the measures described in this paper can be generalised to weighted networks \cite{Ellens2}. Effort should be made to also define robustness measures for \emph{flow networks}; networks for which a \emph{traffic matrix} (with entries that denote the amount of traffic between two vertices) and \emph{edge capacities} (bounds on the load of an edge) are given.

\bibliography{Bibliography}
\bibliographystyle{plain}

\end{document}